\documentclass[reprint,twocolumn,showpacs,amsmath,amssymb,aps,floatfix,superscriptaddress,longbibliography,nofootinbib]{revtex4-2}
\usepackage{amsthm}
\usepackage{amsmath, mathtools}
\usepackage[utf8]{inputenc}
\usepackage[T1]{fontenc}
\usepackage[english]{babel}
\usepackage{amsmath}
\usepackage{graphicx}		
\usepackage{natbib}
\usepackage{textcomp}
\usepackage{gensymb}
\usepackage[usenames,dvipsnames,svgnames,table]{xcolor}
\usepackage[hidelinks,colorlinks=false,urlcolor=Cerulean,citecolor=black]{hyperref}
\usepackage{siunitx}
\usepackage{mathrsfs}
\usepackage{multirow}
\usepackage{bm}
\usepackage{xfrac}
\usepackage{comment}
\usepackage[normalem]{ulem}
\usepackage{enumitem}

\let\dotlessi\i
\renewcommand{\i}{\mathrm{i}}

\begin{document}

\title{Phase-delays shape multistability and basin sizes in Kuramoto networks: \\ analytical estimates from network structure}

\author{Kalel L. Rossi}
\thanks{co-first authors}
\affiliation{Cellular Computations and Learning, Max Planck Institute for Neurobiology of Behavior, Bonn, Germany}

\author{Antonio Mihara}
\thanks{co-first authors}
\affiliation{Department of Physics, Universidade Federal de São Paulo, São Paulo SP, Brazil}

\author{Lyle E. Muller}
\affiliation{Department of Neuroscience, University of Texas at Dallas, Richardson TX, USA}
\affiliation{Fields Lab for Network Computation, Fields Institute, Toronto ON, Canada}

\author{Rene O. Medrano-T}
\affiliation{Department of Physics, Universidade Federal de São Paulo, São Paulo SP, Brazil}
\affiliation{Department of Physics, Universidade Estadual Paulista, São Paulo SP, Brazil}

\author{Roberto C. Budzinski}
\email{roberto.budzinski@uleth.ca}
\affiliation{Department of Neuroscience, University of Lethbridge, Lethbridge AB, Canada}
\affiliation{Fields Lab for Network Computation, Fields Institute, Toronto ON, Canada}

\begin{abstract}
We study how network connectivity and heterogeneous phase-delays shape the spatiotemporal dynamics of finite oscillator networks. Phase-delays can destabilize global synchronization and promote phase-locked patterns, including states with uniform phase gradients and more complex combinations of these modes. Yet, how connectivity and phase-delays jointly determine which states the network selects remains unclear. Here, we show that the spectrum of a composite matrix, which combines connectivity and phase-delays, governs not only the linear stability of the network's collective states but also their basin sizes. This, in turn, enables analytical estimates of basin size of phase-locked states for individual networks from connectivity and phase-delays alone. Applying this framework to nonlocal and global networks, including cases with random phase-delays, we uncover multistability and strong asymmetries in basin sizes, revealing chiral dynamics that conventional stability analysis cannot detect.
\end{abstract} 

\maketitle

\section*{Introduction}

The emergence of organized spatiotemporal dynamics is ubiquitous in natural and engineered systems. Neural systems exhibit structured oscillations and travelling waves linked to function, behaviour, and learning \cite{ermentrout2001traveling, muller2018cortical, muller2016rotating, mohan2024direction}; biological populations display coordinated patterns, as observed in fireflies \cite{buck1976synchronous,buck1988synchronous}, crabs \cite{rorato2017social}, and even human crowds \cite{gu2025emergence}; and power grids rely on coherent dynamical operation for stability \cite{motter2013spontaneous,tyloo2019key}. A central challenge is to understand how the structure of interactions shapes the emergence of such spatiotemporal patterns and, ultimately, system-level function. While significant progress has been made in characterizing phase synchronization \cite{rosenblum1996phase,pecora1998master,arenas2008synchronization,bayani2024transition}, many systems exhibit a richer repertoire of collective states, including travelling waves and phase-locked patterns \cite{ermentrout1981n,crook1997role,jeong2002time,ko2007effects,laing2016travelling,ruschel2025master}, and other forms of partial coordination \cite{abrams2004chimera,abrams2008solvable,omel2018mathematics}. Understanding not only which states exist, but also which are stable and thus dynamically accessible, remains a fundamental problem.

Phase oscillator models, particularly Kuramoto-type systems \cite{acebron2005kuramoto,rodrigues2016kuramoto}, provide a canonical framework to study these phenomena. Despite their simplicity, these models display a wide range of dynamical behaviours and have been successfully used to describe both natural and engineered systems \cite{breakspear2010generative, dorfler2014synchronization}. A substantial body of work has focused on the stability of the phase-synchronized state and, more generally, on the coexistence of multiple stable states \cite{wiley2006size,delabays2017size,townsend2020dense,mihara2019stability, mihara2022sparsity, kassabov2021sufficiently}. However, linear stability alone does not determine which states are observed. In multistable systems, the outcome depends on the basins of attraction, which are difficult to characterize. With this, predicting the likelihood of different collective states often relies on extensive numerical simulations \cite{datseris2022effortless, datseris2023framework}. Even in relatively simple networks, determining both stability and basin structure remains challenging \cite{wiley2006size,delabays2017size,zhang2021basins,groisman2025size,yan2026basin}.

Beyond network connectivity, the form of the coupling itself plays a crucial role in shaping the dynamics. In particular, phase-delays in the interaction between oscillators can drastically alter the dynamical landscape and linear stability properties, enabling the emergence of travelling waves \cite{budzinski2023analytical,an2024stability,lee2024stability,sinha2025geometric}, complex spatiotemporal patterns \cite{shanahan2010metastable,wolfrum2011chimera,omel2012stationary,omel2012nonuniversal}. However, a general mathematical framework that captures how network connectivity and heterogeneous phase-delays jointly determine both stability and basin organization is still lacking.

Here, we introduce a spectral framework that addresses this gap, building on a composite matrix that encodes both network connectivity and coupling phase-delays in oscillator networks \cite{budzinski2022geometry,budzinski2023analytical}. With this, we show here that the eigenvalues of this matrix determine not only the linear stability of phase-synchronized and phase-locked states, but also the size of their basins of attraction, even in the presence of multiple, heterogeneous phase-delays. This allows us to estimate basin sizes directly from network structure, without resorting to extensive numerical simulations. We apply this framework to networks with nonlocal coupling and heterogeneous phase-delays, where we uncover multistability among distinct phase-locked states and analytically estimate their basin sizes; among these, we identify multi-phase-locked states, which combine fundamental phase-locked patterns and broaden the known repertoire of collective behaviour in these systems. Lastly, we apply this framework to globally coupled networks with structured and random phase-delays, showing that phase-delays can reorganize stability, promote phase-locked states, and induce strong asymmetries in basin sizes, revealing signatures of chiral dynamics.

\section*{Oscillator networks, spectrum and dynamics}

We consider networks of nonlinear oscillators defined on a graph with adjacency matrix $\bm{A}$, whose dynamics is given by the Kuramoto-Sakaguchi model:
\begin{equation}
    \dot{\theta}_{i}(t) = \epsilon \sum_{j=0}^{N-1} A_{ij} \sin{\big(\theta_{j}(t) - \theta_{i}(t) -\phi_{ij} \big)},
    \label{eq:main_kuramoto}
\end{equation}
where $\theta_{i}(t)$ is the phase of oscillator $i$ at time $t$, $N$ is the number of oscillators, $\epsilon$ is a scalar factor on the coupling strength that does not alter the generality of the results, $A_{ij}$ are the elements of the adjacency matrix $\bm{A}$, and $\phi_{ij}$ is the phase-delay (or phase-lag) in the interaction between nodes $i$ and $j$. We note that the phase-delays can approximate time-delays in the interactions between oscillators \cite{yeung1999time,ko2007effects,budzinski2023analytical}.

Our main goal is to determine how the network connectivity and phase-delays in the coupling shape the resulting dynamics of the system. To do so, we leverage an complex-valued framework of the Kuramoto model \cite{budzinski2022geometry,budzinski2023analytical} that expresses the dynamics of these networks in terms of operators, and with that, introduces a composite, complex-valued matrix that incorporates connectivity and phase-delays, whose elements are given by
\begin{equation}
K_{ij} = \epsilon e^{-\i \phi_{ij}} A_{ij}.
\label{eq:composite_matrix}
\end{equation}
The spectrum of this matrix predicts and explains the nonlinear dynamics of Eq.~\eqref{eq:main_kuramoto}. Specifically, the phase (or argument) of the elements of the eigenvectors of $\bm{K}$ are possible solutions of the system \cite{budzinski2022geometry,budzinski2023analytical,nguyen2023equilibria}, and the eigenvalues of $\bm{K}$ reveal their linear stability \cite{sinha2025geometric}. This correspondence follows a complex-valued formulation of the Kuramoto model \cite{muller2021algebraic} that supports an operator description of the system \cite{budzinski2022geometry,budzinski2023analytical}, which expresses its dynamics as a combination of a nonlinear and a linear operator applied iteratively. Here, we use the spectrum of $\bm{K}$ to predict and interpret the resulting stability and basin structure of Kuramoto networks, while all the dynamics are obtained directly from the numerical integration of Eq.~\eqref{eq:main_kuramoto}, which are used to confirm the predictions. We note that $ \i = \sqrt{-1}$ is the imaginary unit and $i$ is an index. Details on the operator description of the Kuramoto model and on the eigenspectrum of $\bm{K}$ can be found in Methods Secs.~\ref{sec:operator_KM} and \ref{sec:cdt}.

\section*{Distance-dependent phase-delays shape basins size}

\begin{figure*}[htb]
    \centering
    \includegraphics[width=\linewidth]{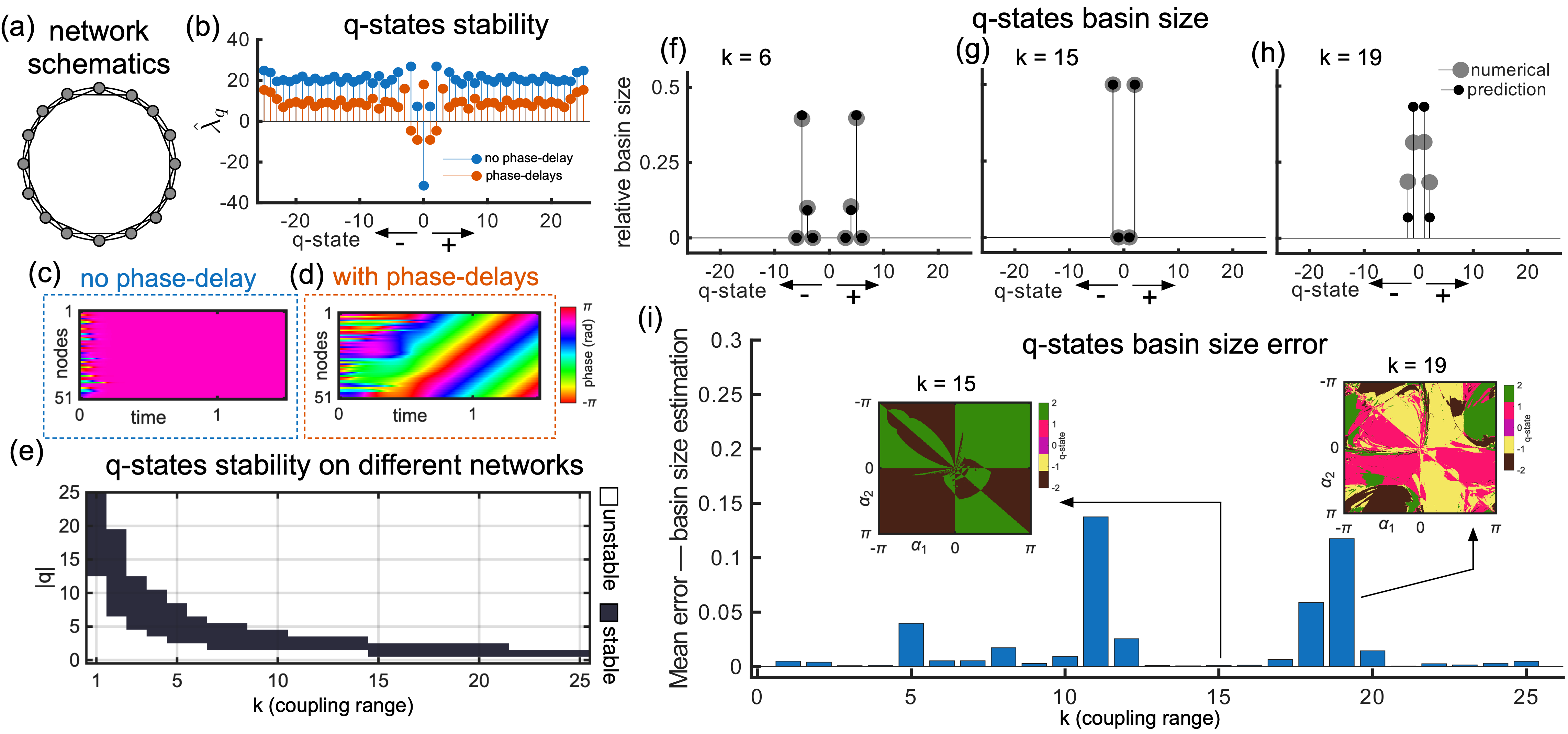}
    \caption{\textbf{Network connectivity and phase-delays change stability and basin sizes of $q$-states.} \textbf{(a)} We consider networks with nonlocal coupling, where each node is connected to $k$ neighbours on each side on the ring ($k$-ring graphs). \textbf{(b)} Linear stability of $q$-states derived from the eigenvalues of $\bm{K}$. Here, $\hat{\lambda}_q = \mathrm{max}(\lambda_{m,q})$ indicates the maximum value for a given $q$ over all directions $m$. If $\hat{\lambda}_q < 0$, then the $q$-state is linearly stable. Blue (orange) dots indicate stable states in the absence (presence) of phase-delays. Here we consider a network with $k = 20$ and $N = 51$. \textbf{(c-d)} Spatiotemporal patterns observed in the system: without phase-delays only the phase synchronized state is stable, whereas distance-dependent phase-delays stabilize phase-locked states. \textbf{(e)} Stability diagram as a function of coupling range $k$, showing the emergence of different stable $q$-states and the destabilization of the synchronized state across all coupling range in the presence of distance-dependent phase-delays. \textbf{(f–h)} Basin sizes of stable $q$-states predicted analytically from the spectrum of $\bm{K}$ (black dots) compared with $20,000$ numerical simulations of Eq.~\eqref{eq:main_kuramoto} with random initial conditions (grey dots), for representative nonlocal coupling ranges. Symmetric pairs $\pm q$ exhibit equal basin sizes. \textbf{(i)} Relative error between analytical predictions and numerical basin sizes (over $20,000$ simulations) across coupling ranges, demonstrating good quantitative agreement and correct ranking of basin sizes even in cases with deviations. Insets display sections of the basin of attraction for the case of $k = 15$ and $k = 19$. We start the system on the phase synchronized state $[0, 0, 0, \cdots, 0]$ (which is unstable), and perturb the first two oscillators with $\alpha_1$ and $\alpha_2$, respectively: $\bm{\theta}_0 = [\alpha_1, \alpha_2, 0, \cdots, 0]$ and measure the final state of the simulation (represented in colour-code).}
    \label{fig:phase_lag_basins}
\end{figure*}

To explore how connectivity and phase-delays affect the spatiotemporal dynamics of oscillator systems, we consider networks with nonlocal coupling, where each node is connected to $2k$ neighbours ($k$ on each side) and the network is arranged in a one dimensional ring (periodic boundary conditions) (Fig.~\ref{fig:phase_lag_basins}a). We also consider distance-dependent phase-delays in the coupling, where the more distant two nodes are, the larger is the phase-delay (see Methods Sec.~\ref{sec:network_phase_lags} for details on the network structure). In this case, both the connectivity and phase-delays matrices are symmetric and circulant \cite{davis1979}, so that $q$-states are solutions to the system: $\bm{\theta}^{(q)} = \big(0, 2\pi q{N}^{-1}, \cdots, 2\pi q(N-1){N}^{-1} \big)$, where $q=0$ represents the phase synchronized state and $|q| > 0$ represent phase-locked states with different spatial frequency. Importantly, the argument of the elements of the eigenvectors of $\bm{K}$ in these cases are exactly $\bm{\theta}^{(q)}$ \cite{budzinski2023analytical}, and the real part $\gamma_j$ of the $j$-th eigenvalue of $\bm{K}$ determines the linear stability of these states \cite{sinha2025geometric}. In particular, the linear stability of a $q$-state along the $m$-th eigendirection of $\bm{K}$ is controlled by:
\begin{equation}
   \lambda_{m,q} = \frac{\gamma_{q+m} + \gamma_{q-m}}{2} - \gamma_{q},
    \label{eq:eigenvalues_cdt_stability}
\end{equation}
so $\bm{\theta}_q$ is linearly stable if $\lambda_{m,q} < 0, \, \forall \, m \in [1,N-1]$ and linearly unstable otherwise.

We first consider a network with $N = 51$ oscillators coupled each with their $k = 20$ nearest neighbours following Eq.~\eqref{eq:main_kuramoto}. In this case, without phase-delays, only the phase synchronized state is attracting, corresponding to a stable fixed point (blue dots, Fig.~\ref{fig:phase_lag_basins}b, and Fig.~\ref{fig:phase_lag_basins}c). If we introduce distance-dependent phase-delays -- Eq.~\eqref{eq:ddphaselag} -- the phase synchronized state loses stability and phase-locked solutions become stable (orange dots, Fig.~\ref{fig:phase_lag_basins}b, and Fig.~\ref{fig:phase_lag_basins}d), corresponding to stable limit cycles, which can be mapped onto fixed points by shifting to a co-moving reference frame. We can change the range of the nonlocal coupling (the number of nearest neighbours $k$), which then leads to the emergence of different $q$-states (Fig.~\ref{fig:phase_lag_basins}e).

In this work, we show that, in addition to the linear stability of these states, the spectrum of $\bm{K}$ also encodes information about the size of their basins of attraction. This starts from recent work showing that the relative basin size of $q$-states in $k$-ring graphs can be estimated from local eigenvalues of the Jacobian \cite{mihara2022basin}. Specifically, from the ``equilibrium stability" of a given $q$-state, given by the sum of all negative eigenvalues of the respective Jacobian, normalized by the most negative eigenvalue, and then raised to the power of the dimension $N$ of the system. Here, we found a direct relationship between the sum of the negative eigenvalues of the Jacobian and the real part of the eigenvalues of the matrix $\bm{K}$ (see Methods Sec.~\ref{sec:basin_size} for details). This connection allowed us to generalize the idea to analytically estimate the relative basin size $\mathscr{S}_{B_q}$ of a stable $q$-state through the estimator $\Gamma_q$:
\begin{equation}\label{eq:basin_size_main}
   \mathscr{S}_{B_q} \approx \Gamma_q = \frac{\gamma_q^N}{\sum_{q\in\mathcal{Q}} \gamma_q^N},
\end{equation}
with $\mathcal{Q}$ being the set of stable $q$-states. 

Because the matrix $\bm{K}$ incorporates network connectivity and phase-delays, our framework now opens a new path to analytically estimate the basin sizes of $q$-states in a diversity of systems, including networks with multiple delays and phase-delays. Using our framework, we are able to (1) analytically determine which $q$-states are linearly stable and (2) estimate the basin of attraction sizes for each of these stable states. For example, we consider the nonlocal networks with distance-dependent phase-delays with different coupling ranges, and using only the eigenvalues of $\bm{K}$ we obtain the basin sizes of the $q$-states in these networks (black dots, Figs~\ref{fig:phase_lag_basins}f-h). We verify these results numerically, where we integrate Eq.~\eqref{eq:main_kuramoto} starting the system with random initial conditions and measure the final state, which then allow us to numerically estimate the relative size of the basin of each of the $q$-states (grey dots, Figs.~\ref{fig:phase_lag_basins}f-h). We note the $q$-states appear in pairs, with positive and negative $q$-states, which represent phase-locking states with the same spatial frequency but different direction of phase offset around the ring \cite{budzinski2023analytical,sinha2025geometric}, and that these solutions are symmetric and display virtually the same basin size, which is captured by our approach.
\begin{figure*}[t!]
    \centering
    \includegraphics[width=0.95\linewidth]{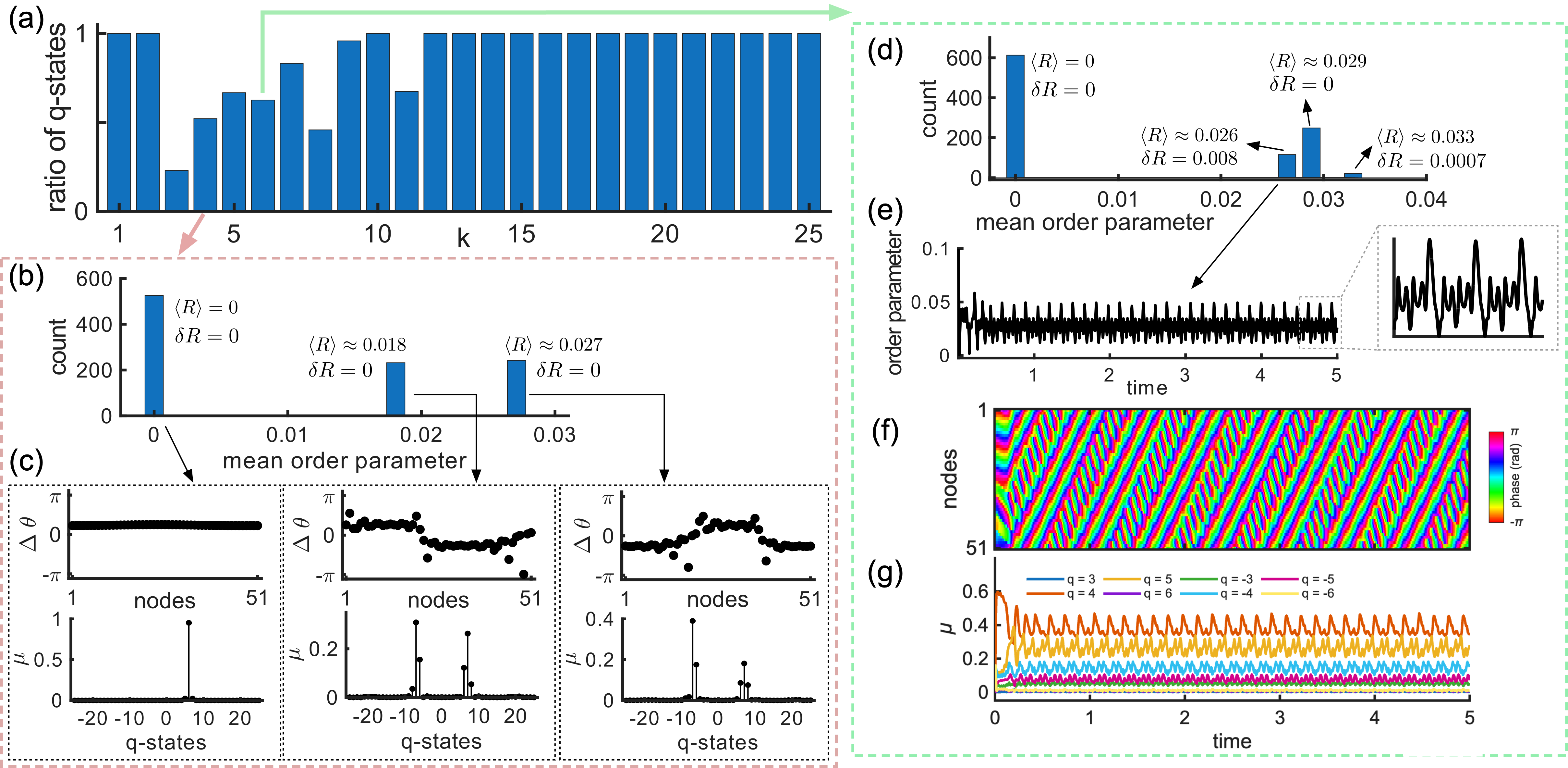}
    \caption{\textbf{Emergence of multi-$q$ states in nonlocally coupled oscillator networks with distance-dependent phase-delays.} \textbf{(a)} Fraction of simulations (with random initial conditions) converging to single $q$-states as a function of coupling range $k$, revealing a regime $k  \in [3,\,11]$ where additional solutions emerge. \textbf{(b)} Example for $k = 4$: histogram of the time-averaged order parameter $\langle R \rangle$ shows three distinct states reached by $\mathrm{count}$ number of initial conditions (here, all three states do not display temporal variations, i.e. $\delta R = 0$). \textbf{(c)} We characterize these states via nearest-neighbour phase differences $\Delta \theta$ and projections $\mu$ onto the eigenmodes of $\bm{K}$ -- cf. Eq.~\eqref{eq:modes_contribution}. The case $\langle R \rangle = 0$ corresponds to pure $q$-states (single eigenmode), while intermediate values ($\langle R \rangle \approx 0.018$ and $\langle R \rangle \approx 0.027$) reveal multi-q states with spatially localized phase locking and contributions from multiple eigenmodes. \textbf{(d)} Example for $k = 6$: histogram of the time-averaged order parameter $\langle R \rangle$ shows 4 different states. In addition to $q = 0$ states ($\langle R \rangle = 0$) and multi-$q$-states ($\langle R \rangle \approx 0.029$) with no temporal variation ($\delta R = 0$), two additional states with fluctuating order parameter ($\langle R \rangle \approx 0.026$ $\delta R= 0.008$, and $\langle R \rangle \approx 0.033$  $\delta R = 0.0007$) are observed. \textbf{(e)} We plot the order parameter over time one of these states, which reveals the temporal fluctuations. \textbf{(f)} The corresponding spatiotemporal evolution of this state, highlights the multi-$q$-state structure with time-dependent modulation, \textbf{(g)} which is confirmed by the temporal evolution of the eigenmodes contributions. Distance-dependent phase-delays follow Eq.~\eqref{eq:ddphaselag}.}
    \label{fig:multi_q_nonstationary}
\end{figure*}

We then systematically analyze the difference between our predictions and the numerical results regarding the basin sizes of $q$-states in these networks. We vary the network connectivity and phase-delays from first neighbours to global networks and measure the error in our predictions (Fig.~\ref{fig:phase_lag_basins}i) -- see Methods Sec.~\ref{sec:error_basin} for details. We note that our approach produces highly accurate predictions for the entire range of networks analyzed here, with a few exceptions where the error increases with a maximum around 13\% in the estimative of the basin size. We note that the cases with increased error are associated with highly complicated basin geometries (see insets, Fig.~\ref{fig:phase_lag_basins}i). Importantly, however, even in these cases, our approach correctly predicts the order of the basin sizes among the $q$-states (see example in Fig.~\ref{fig:phase_lag_basins}h).

\section*{Multi-modal states in networks with distance-dependent phase-delays}

The $q$-states are not the only possible solutions we observed in these networks. In fact, for specific values of $k$ (coupling range), different solutions appear. We integrate Eq.~\eqref{eq:main_kuramoto} with distance-dependent phase-delays -- Eq.~\eqref{eq:ddphaselag} -- starting the system with many different randomly chosen initial conditions and measure the ratio of trajectories that reach a $q$-state (Fig.~\ref{fig:multi_q_nonstationary}a). For a specific range of $k \in [3,\,11]$ we observe other states than $q$-states emerging in the system. To exemplify some of these possible states we focus on two cases. To evaluate the different states the network can reach, we calculate the order parameter $R(t) = \big| \sfrac{1}{N} \sum_{j=0}^{N-1} e^{\i \theta_j(t)} \big|$, which measures the level of phase synchronization. For phase synchronized states (($q=0$)-state), $R = 1$, and for phase-locked states ($(|q|>0)$-states), $R = 0$. Other states display intermediate values of $R$. Here, we calculate the average $\langle R \rangle$ and the standard deviation $\delta R$ of the order parameter over time after discarding an initial transient.

To explore different examples of these possible states, we first consider the case of $k = 4$, where we observe three possible values of the mean order parameter $\langle R \rangle$ (Fig.~\ref{fig:multi_q_nonstationary}b, $\mathrm{count}$ indicates number of initial conditions converging to each value).  All trajectories reach stable limit cycles. To study these states, we evaluate the phase difference between oscillators $\Delta \theta_{i} = \theta_{i+1} - \theta_i$ (with respect to the periodic boundaries conditions) and also the projection of the dynamical state at the end of the simulation $\bm{\theta}$ onto the eigenvectors of $\bm{K}$ (eigenmodes) (see Methods Sec.~\ref{sec:eigenmodes} for details). The first type of state is a $q$-state, for which $\langle R \rangle = 0$  and the phase difference between oscillators is constant, and only one eigenmode is observed (corresponding to the $q$-state itself) (left, Fig.~\ref{fig:multi_q_nonstationary}c). The second type of state is characterized by $\langle R \rangle \approx 0.018$ and has two clusters of oscillators with roughly similar phase differences. The eigenmode decomposition reveals the contribution of different $q$-states (middle, Fig.~\ref{fig:multi_q_nonstationary}c). A similar type of state is observed for the third case, with $\langle R \rangle \approx 0.027$. In this case, we observe a specific configuration of phase difference between oscillators and a set of eigenmodes contributing to the dynamics. These states are very similar to multi-$q$-states, or ``multi-twisted'' states, first observed in nonlinear oscillator networks with repelling coupling \cite{girnyk2012multistability}.

As a second example of the richness of the dynamics emerging in these systems, for $k = 6$, we observe different values of $\langle R \rangle$ (Fig.~\ref{fig:multi_q_nonstationary}d). For $\langle R \rangle = 0$, the system reached $q$-states. We also observed a multi-$q$-state with $\langle R \rangle \approx 0.029$. For these networks, however, we also observed two additional states where the order parameter asymptotically is not constant over time ($\langle R \rangle \approx 0.026$ $\delta R= 0.008$, and $\langle R \rangle \approx 0.033$  $\delta R = 0.0007$, Fig.~\ref{fig:multi_q_nonstationary}d). These states have also similarities with multi-$q$-states, but they are no longer displaying patterns that are static over time. As an example, we consider the state with $\langle R \rangle \approx 0.026$, which shows temporal variations of $R(t)$ (Fig.~\ref{fig:multi_q_nonstationary}e). When we analyze the spatiotemporal dynamics, it becomes clear the multi-$q$-state nature of this state and the temporal variation (Fig.~\ref{fig:multi_q_nonstationary}f). In fact, we observe that eigenmodes' contribution also vary over time, as the network changes its spatiotemporal pattern (Fig.~\ref{fig:multi_q_nonstationary}g).

\section*{Chiral dynamics in globally coupled network with random phase-delays}

For globally connected networks without phase-delays, the phase synchronized state is the only stable solution. However, as we showed previously, when we add distance-dependent phase-delays, the phase synchronized state loses stability and a pair of $q$-states appear instead (Fig.~\ref{fig:phase_lag_basins}). These two state have the same spatial frequency, but different phase offset direction, and their basins of attraction have the same relative size. With this, random initial conditions have equal probability to transition to any of these two state. This is usually the case in symmetric networks, as these $q$-states appear in pairs. 
\begin{figure}[thb]
    \centering
    \includegraphics[width=\linewidth]{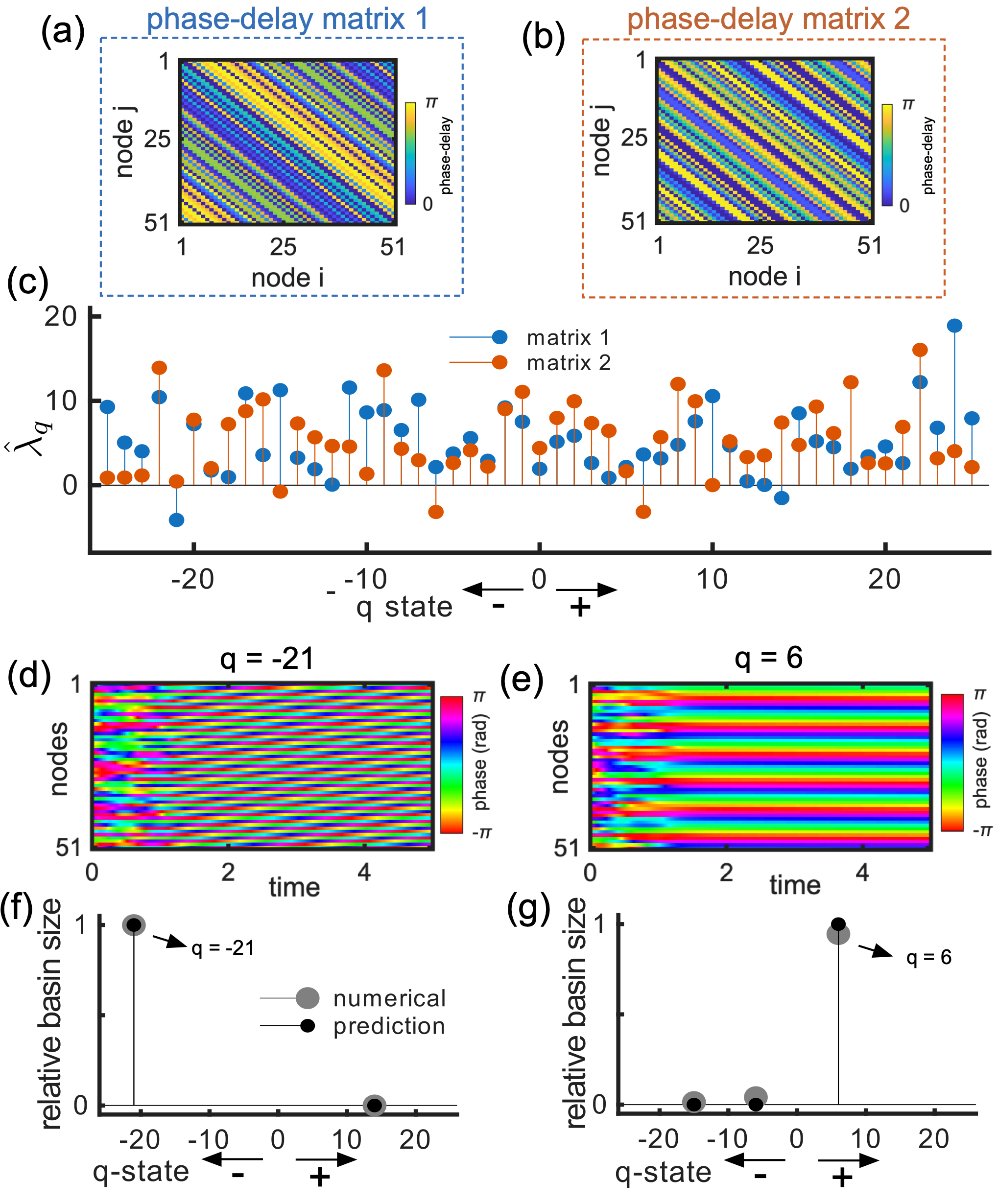}
    \caption{\textbf{Stability and basin asymmetry of $q$-states in globally coupled networks with random phase-delays.} \textbf{(a–b)} We consider global networks with random and circulant phase-delays (matrices 1 and 2), which preserve the existence of $q$-state solutions. \textbf{(c)} Linear stability analysis from the spectrum of $\bm{K}$ -- given by $\hat{\lambda}_q$ -- identifying stable $q$-states: for case 1, $q = -21$ and $q = 13$; for case 2, $q = -15$, $q = -6$, and $q = 6$. \textbf{(d–e)} Representative simulations starting with random initial conditions showing convergence to different stable $q$-states. \textbf{(f–g)} Analytical estimation of basin sizes (black dots) based on the eigenvalues of $\bm{K}$ for both cases. We perform numerical simulations (starting at random initial conditions) to obtain the basin size (grey dots), which confirm the analytical predictions. Unlike the distance-dependent phase-delay cases, stable states do not necessarily occur in symmetric $\pm q$ pairs, and even when they do (e.g., $q = \pm 6$ in case 2), their basin sizes can differ significantly, indicating the emergence of chiral dynamics despite global connectivity.}
    \label{fig:global_random}
\end{figure}

Here, we now consider random, but circulant, distributed phase-delays (Figs.~\ref{fig:global_random}a, \ref{fig:global_random}b). In these cases, despite being random, the matrices representing the phase-delays are still circulant, and thus the $q$-states are still possible solutions to the system \cite{sinha2025geometric}. We then use our framework to obtain the stability of these states (given by $\hat{\lambda}_q$), and we observe that for case 1, $q = -21$ and $q = 13$ state are stable, and for case 2, $q = -15$, $q = -6$, and $q = 6$ states are stable (Fig.~\ref{fig:global_random}c). To exemplify some of these cases, we simulate the network starting on random initial conditions, where we can observe the emergence of different $q$-states (Figs.~\ref{fig:global_random}d, \ref{fig:global_random}e). Further, we can now analytically estimate the basin size of these states using our approach (black dots), which is then confirmed by numerical simulations (grey dots) in Figs.~\ref{fig:global_random}f and \ref{fig:global_random}g. We note this reflects a different scenario than observed before: we no longer have only pairs of $q$-states being stable, and even in the case 2, where the pair of $q$-states ($q = -6$, and $q = 6$) have similar stability properties (orange dots, Fig.~\ref{fig:global_random}c), their basins are very different (Fig.~\ref{fig:global_random}g), which is analytically captured by our approach. This represents a case where even global connectivity can lead to asymmetric and chiral dynamics.

\section*{Discussion}

We investigated how network connectivity and heterogeneous phase-delays shape the stability landscape and basin structure of nonlinear oscillator networks. We leveraged a composite matrix $\bm{K}$ that incorporates both connectivity and phase-delays, and showed that its eigenvalues provide a unified spectral framework linking the linear stability and basin sizes of $q$-states directly to the network's structure. The resulting framework provides an analytical route to predict not only which collective states are stable, but also to estimate their relative likelihood of emergence from random initial conditions.

For nonlocal networks with structured, distance-dependent phase-delays, we showed that phase synchronization is destabilized and replaced by phase-locked $q$-states whose stability depends on the nonlocal coupling range. Importantly, we demonstrated that the same spectral information used for linear stability also yields accurate analytical estimates of basin sizes. Determining the stability and basins properties of networked systems has been an important question in nonlinear dynamics and complex systems, and even in simple networks like in the case of coupled oscillator systems, it remains challenging. It has been shown in simple cases without phase-delay that the coupling range can change the stability of $q$-states \cite{wiley2006size,mihara2019stability}, and the properties of their basins has been a topic of debate \cite{mihara2022basin,zhang2021basins,groisman2025size}. Recent work has explored the geometric properties of these basins in networks with nearest-neighbour coupling and shown their rich complexity \cite{zhang2021basins}. In this context, recent work has introduced an analytical form to estimate their basin size based on the local eigenvalues of stable states in simple cases without phase-delays \cite{mihara2022basin}. Here, we have now pushed this forward and based on our framework, we are able to determine the stability and basin properties even in the presence of multiple phase-delays in the coupling. Across a wide range of parameters, our predictions closely match numerical simulations and correctly capture the ordering of basin sizes, even in cases where quantitative discrepancies arise.

Beyond $q$-states, we uncovered a richer dynamical repertoire, including static and time-dependent multi-$q$ states. These states are characterized by contributions from multiple eigenmodes. Their emergence highlights that the dynamics cannot be fully understood from linear stability alone, and that nonlinear interactions between modes play a key role in shaping the behaviour of the system.

Finally, by considering random (but circulant) phase-delays, we demonstrated that even in globally coupled and structurally homogeneous networks, the basin structure can become highly asymmetric. There has been recent interest in studying the dynamics of oscillator networks with multiple phase-delays \cite{lee2024stability,an2024stability,sinha2025geometric}, which can have important applications in neuroscience \cite{koller2024human,budzinski2023analytical,dugue2025traveling} and neural computation \cite{benigno2023waves, ricci2021kuranet}. Recent work has shown that the collective dynamics of global networks with multiple phase-shifts in the coupling can be expressed in low-dimensional representation of the system \cite{smirnov2024dynamics}. However, this approach is studied in the case of monostable system, and analyzing the properties of multistable systems remains challenging. To this end, we have shown that our approach can be in fact applied to cases where the system is multistable. In particular, we observed cases where pairs of $q$-states with similar linear stability exhibit markedly different basin sizes, revealing an effective chiral symmetry breaking in the dynamics. This result emphasizes that basin structure provides complementary and essential information beyond stability analysis.

Overall, our work establishes a general and tractable framework to study stability and basin organization in oscillator networks with heterogeneous interactions. These findings open new directions for controlling collective dynamics through network connectivity and coupling functions and suggest that spectral properties can serve as powerful predictors of emergent behaviour in complex dynamical systems.

\section{Methods}

\subsection{Operator description of the Kuramoto model}\label{sec:operator_KM}

The Kuramoto model (KM) -- Eq.~\eqref{eq:main_kuramoto} -- can be understood by analyzing a closely related dynamical system defined in the complex numbers \cite{muller2021algebraic}. Specifically, the KM can be re-expressed in terms of a recursion with operators -- one linear, and one nonlinear -- applied iteratively \cite{budzinski2022geometry,budzinski2023analytical}. This recursion recovers the precise trajectory in individual realizations of the Kuramoto system, starting from arbitrary initial conditions. This operator description produces the same dynamical trajectory as the original Kuramoto system; however, the utility of the approach is that it allows making analytical predictions from the operator expression, by eigen-expanding under the nonlinearity. Here, we summarize the derivation of this approach, starting from the exact solution of the complex-valued system \cite{muller2021algebraic} to the operator description and analytical predictions \cite{budzinski2022geometry,budzinski2023analytical}.

Consider a complex-valued vector state $\bm{\psi} \in \mathbb{C}^N$ with the network's dynamics given by:
\begin{equation}
    \begin{split}
        \dot{\psi}_{i}(t) = \epsilon\sum_{j = 1}^{N} a_{ij}\Big( \sin{ \big(\psi_{j}(t) - \psi_i(t) - \phi_{ij} \big)} \\ -\i \cos{ \big(\psi_{j}(t) - \psi_i(t) - \phi_{ij} \big)} \Big),
    \end{split}
    \label{eq:nonlinear}
\end{equation} 
where $\epsilon$, $A_{ij}$, and $\phi_{ij}$ are the same as in the Kuramoto model Eq.~\eqref{eq:main_kuramoto}. We can then multiply Eq.~\eqref{eq:nonlinear} by $\i$, use Euler's formula, and rearrange the exponential terms, which leads to:
\begin{equation}
     \i \dot{\psi}_i(t)= \epsilon \sum_{j=1}^N a_{ij} e^{\i \psi_j(t)} e^{-\i \psi_i(t)}  e^{-\i \phi_{ij}},
      \label{eq:aux_psi_1}
\end{equation}
or
\begin{equation}
 \i e^{\i \psi_i(t)}  \dot{\psi}_i(t)= \epsilon \sum_{j=1}^N a_{ij} e^{-\i \phi_{ij}} e^{\i \psi_j(t)}.
 \label{eq:aux_psi_2}
\end{equation}
We can then use a nonlinear change of variable, such that $x_i (t) =e^{\i \psi_i(t)}$, where $\dot{x}_{i}(t) = \i e^{\i \psi_i(t)}  \dot{\psi}_i(t)$. With this, we can re-express the equation above as:
\begin{equation} 
\dot{x}_{i}(t) = \epsilon \sum_{j=1}^N a_{ij} e^{-\i \phi_{ij}} x_{j}(t).
\label{eq:linear_cv_eq}
\end{equation}
From here, we can introduce a composite matrix $\bm{K}$, whose elements are
\begin{equation}
    K_{ij} = \epsilon e^{-\i \phi_{ij}} a_{ij}.
\end{equation}
We note Eq.~\eqref{eq:linear_cv_eq} has a closed-form solution given by:
\begin{equation}
    \bm{x}(t) = e^{\bm{K} t} \bm{x}(0).
    \label{eq:solution}
\end{equation}
We now note that from the nonlinear change of variables, we have that $\mathrm{Arg}[x_{i}(t)] = \mathrm{Re}(\psi_{i}(t))$. So, we can write Eq.~\eqref{eq:aux_psi_1} explicitly with real and imaginary parts of $\psi$. The real part is given by:
\begin{equation}
    \mathrm{Re}\big(\dot{\psi_i}(t)\big) = \epsilon \sum_{j=1}^N a_{ij} \frac{|x_{j}(t)|}{|x_{i}(t)|} \sin{\Big( \mathrm{Re}\big(\psi_{j}(t)\big) - \mathrm{Re}\big(\psi_{i}(t)\big) - \phi_{ij} \Big)}.
    \label{eq:real_part_psi}
\end{equation}
This shows that Eq.~\eqref{eq:real_part_psi} is exactly Eq.~\eqref{eq:main_kuramoto} when $\frac{|x_{j}(t)|}{|x_{i}(t)|} = 1$. In this case, $\mathrm{Arg}[x_{i}(t)] = \mathrm{Re}(\psi_{i}(t)) = \theta_{i}(t)$. To obtain the operator-description of the Kuramoto model, we then use the solution of $\bm{x}(t)$ in addition to a nonlinear operator $\Upsilon$:
\begin{equation}
    \bm{x}(t+\varsigma) = \Upsilon \big[e^{\varsigma \bm{K}} \bm{x}(t)\big],
    \label{eq:cv_approach}
\end{equation}
where $\varsigma$ is a fixed timestep. We propagate the solution of $\bm{x}(t)$, Eq.~\eqref{eq:solution}, for short timesteps $\varsigma$ using the matrix exponential, then use the nonlinear operator $\Upsilon[x_{i}] = \sfrac{x_{i}}{|x_{i}|}$. This ensures the condition $\frac{|x_{j}(t)|}{|x_{i}(t)|} = 1$, and the equivalence between the original and the operator description of the Kuramoto model. With this, we can link the spectrum of $\bm{K}$ with the emergent dynamics in Kuramoto networks. 

Here, based on the insights provided by this operator-description, we use the spectrum of the composite matrix $\bm{K}$ to obtain analytical insights on the stability and basin of attraction of $q$-states in finite Kuramoto networks. Specifically, Eq.~\eqref{eq:eigenvalues_cdt_stability} gives us the linear stability, and Eq.~\eqref{eq:basin_size} gives us prediction of the basin size of $q$-states in terms of the eigenvalues of $\bm{K}$ in an analytical manner, without the need of simulations. We also note that all these predictions are confirmed by direct numerical simulations of Kuramoto model given by Eq.~\eqref{eq:main_kuramoto}.

\subsection{Eigenvalues and eigenvectors of circulant matrices}\label{sec:cdt}

The eigenvalues and eigenvectors of any circulant matrix are given by the Circulant Diagonalization Theorem (CDT) \cite{davis1979}. The $j^{\mathrm{th}}$ eigenvalue is given by
\begin{equation}
    \varphi_{j} = \sum_{k=0}^{N-1} h_{k} \exp{\left(\frac{-2\pi \i}{N} jk \right)},
    \label{eq:eigenvalue_cdt}
\end{equation}
where $\bm{h}$ is the generating vector for the matrix. The $s$ entry of the $j^{\mathrm{th}}$ eigenvector is given by
\begin{equation}
    (\bm{v}_{j})_s = \frac{1}{\sqrt{N}} \exp{\left( \frac{-2\pi \i}{N} j s \right)}.
\end{equation}
We note that any $N \times N$ circulant matrix is diagonalized by the same unitary matrix. With this, the eigenvectors are the same and the eigenvalues depend on the generating vector. In our approach, the generating vector of $\bm{K}$ is defined by the pattern of connectivity and phase-delays. Further, we note the argument of the eigenvectors represent directly the $q$-states solutions and their stability and basin properties depend on the real part of the eigenvalues $\gamma_j = \mathrm{Re}(\varphi_j)$:
\begin{equation}
    \gamma_j = \sum\limits_{k=0}^{N-1} A_{0k} \cos{\Bigg(\frac{-2\pi}{N} jk - \phi_{0k} \Bigg)}.
    \label{eq:eigenvalues_real}
\end{equation}

\subsection{Network connectivity and phase-delays}\label{sec:network_phase_lags}

In our work, we consider networks given k-ring graphs. These networks can instantiate local and nonlocal coupling, and globally connected networks. The adjacency matrix of a k-ring graph is given by:
\begin{eqnarray*}
    A_{ij} &= 1, \,\, \mathrm{if} \,\, d_{ij} \leq \mathrm{k}, \\
    A_{ij} &= 0, \,\, \mathrm{otherwise},
\end{eqnarray*}
where $d_{ij} = \mathrm{min}(|i - j|, N - |i - j|)$, which ensures the periodic boundary conditions around the ring.

The distance-dependent phase-delays are given by
\begin{equation}\label{eq:ddphaselag}
    \phi_{jk} = \frac{\pi d_{jk}}{\mathrm{k}}.
\end{equation}

For the random, circulant phase-delays, we create a generating vector $\bm{h} = (h_1, h_2, \cdots, h_N)$, where each element is randomly chosen in the interval $[0, \pi]$. After that, we use a circular perturbation operator on the vector $\bm{h}$ to obtain the phase-delay matrix $\bm{\phi}$. Because of this, the row sum of $\bm{\phi}$ is the same for all rows.

\subsection{Equilibrium stability and basin sizes}\label{sec:basin_size}

The concept of equilibrium stability was introduced in \cite{mihara2022basin}, which depends on the sum of the negative eigenvalues of the Jacobian of a given $q$-state. In Ref.~\cite{mihara2022basin} the stability measure $\hat{\Lambda}_q$ was defined from the sum of (real part of) eigenvalues of the Jacobian of Eq.~\eqref{eq:main_kuramoto} for k-rings with $\phi_{ij} = 0$. First, one takes $\Lambda_q = \sum_m \lambda_{m,q}$ for stable $q$-states. We note that the fully synchronized state ($q=0$) has the most negative value of $\Lambda$ and also the larger basin of attraction for these networks \cite{mihara2022basin}. The equilibrium stability measure was then defined as $\hat{\Lambda}_q = \Lambda_q / \Lambda_{(q=0)}$. Further, as shown in Ref.~\cite{mihara2022basin}, $\hat{\Lambda}_q^N$ can be approximated by a Gaussian distribution with standard deviation $\sim 0.2\sqrt{N/\mathrm{k}}$, in accordance with the results of previous results on the basins of attraction of $q$-states in k-ring graphs Ref.~\cite{wiley2006size}. This can be obtained analytically, where the sum of the negative eigenvalues of the Jacobian is given by \cite{mihara2022basin}. For the systems considered here given by Eq.~\eqref{eq:main_kuramoto} with $\phi_{ij} = 0$, this is given by:
\begin{equation}
    \Lambda_q \approx \Bigg(1 - \frac{\sin{\big((2\mathrm{k}+1)q\pi N^{-1}\big)}}{\sin{(q\pi N^{-1})}}\Bigg)N.
\end{equation}
Importantly, this is directly related to the real part of the eigenvalues of the matrix $\bm{K}$ under the same conditions, which are given by Eq.~\eqref{eq:eigenvalues_real}, and can explicitly be written as \cite{sinha2025geometric}:
\begin{equation}
   \gamma_q =  \frac{\sin{\big((2\mathrm{k}+1)q\pi N^{-1}\big)}}{\sin{(q\pi N^{-1})}} - 1.
\end{equation}
This, in turn, shows that the sum of the negative eigenvalues of the Jacobian can be directly expressed in terms of the eigenvalues of $\bm{K}$, where
\begin{equation}\label{eq:relation_sum_eigenvalues_mihara}
    \gamma_q \approx \frac{-\Lambda_{q}}{N}.
\end{equation}
With this, we are able to analytically estimate the relative basin size $\mathscr{S}_{B_q}$ of a given stable $q$-state in the network of interest. To do so, we first  obtain the real part $\gamma_q$ of the eigenvalues of $\bm{K}$, given by Eq.~\eqref{eq:eigenvalues_real}. With this, we can use Eq.~\eqref{eq:eigenvalues_cdt_stability} to determine the linear stability of the $q$-states, and collect the stable states in the set $\mathcal{Q}$. For $q \in \mathcal{Q}$, we follow Ref.~\cite{mihara2022basin} combined with the insight from Eq.~\eqref{eq:relation_sum_eigenvalues_mihara} to compute the measure $\Gamma_q$ that estimates the relative size $\mathscr{S}_{B_q}$ of the basin for each stable $q$-state:
\begin{equation}\label{eq:basin_size}
   \mathscr{S}_{B_q} \approx \Gamma_q = \frac{\gamma_q^N}{\sum_{q\in\mathcal{Q}} \gamma_q^N}.
\end{equation}

We note that the equilibrium stability introduced in \cite{mihara2022basin} was defined between zero and one, such that basin size distribution of $q$-states in $k$-ring graphs follow a Gaussian distribution. Here, we normalize $\Gamma_q$ such that the $\sum_q \Gamma_q = 1$ over all stable $q$-states, and $\Gamma_q$ is a direct estimate of the relative size of the basin of the respective $q$-state.

\subsection{Error calculation of the basin size estimation}\label{sec:error_basin}

To numerically estimate the basin sizes of $q$-states and confirm our predictions, we numerically integrate Eq.~\eqref{eq:main_kuramoto} staring the system with random initial conditions $\bm{\theta}_0 \in [-\pi, \pi]$. After $t = 10,000$ timesteps, we then calculate a similarity measurement
\begin{equation}
    \mathcal{S}^{(q)} = \frac{1}{N}\left|\sum_{j=0}^{N-1} \exp{\Big(\i \theta_{j}- \i\theta_{j}^{(q)}\Big)}\right|
    \label{eq:similarity}
\end{equation}
between the network state $\bm{\theta}$ at the end of the simulation and the $q$-state $\bm{\theta}^{(q)}$, where $\mathcal{S}^{(q)} = 1$ indicates the network is at the state $q$-state. We then perform this calculation for all simulations and count the number of simulations that reach a given $q$-state. We then extend this over all stable $q$-state and then normalized the quantifier by the total number of simulations that reached $q$-states. This allows us to obtain the numerical estimation of the relative basin size of each stable $q$-state $\mathscr{S}_{B_q}$.

The analytical prediction $\Gamma_q$ and the numerical estimation $\mathscr{S}_{B_q}$ for the basin size are given by a real number between 0 and 1. The error between the prediction and the numerical results for a specific network is given by:
\begin{equation}
    \mathcal{E} = \frac{1}{M}\sum_{q \in \mathcal{Q}} \Big| \mathscr{S}_{B_q} - \Gamma_{q} \Big|,
    \label{eq:error_basin}
\end{equation}
where the sum considers only the stable $q$-states and $M$ is the total number of stable $q$-states. With this, $\mathcal{E} = 0$ means a perfect agreement between the prediction and the numerical estimation of the basin size.

\subsection{Network's eigenmodes}\label{sec:eigenmodes}

We calculate the contribution of each eigenmode of the system on the spatiotemporal dynamics by using the eigenvectors of $\bm{K}$. Specifically, we calculate the complex inner product:
\begin{equation}
    \nu_q(t) = \langle e^{\i\bm{\theta}(t)}, \bm{v}_q \rangle,
\end{equation}
where $\bm{\theta}(t)$ is the state of the system at time $t$ and $\bm{v}_q$ is the $q$-th eigenvector of $\bm{K}$.

We then calculate
\begin{equation}
    \mu_q(t) = \frac{|\nu_q(t)|^{2}}{N},
    \label{eq:modes_contribution}
\end{equation}
which gives us the relative contribution of mode $q$ at time $t$. With this, $\mu_q(t) = 1$ means only mode $q$ has a contribution to the dynamics at time $t$.

\subsection{Step-by-step description of the analyses}\label{sec:details_analysese}

For clarity, we describe below the step-by-step process to obtain the analytical estimates of the basin sizes and for the analyses performed in the paper.
\begin{enumerate}[noitemsep]
    \item Compute eigenvalues of $\bm{K}$ and take their real parts to get $\gamma_i$ using the closed-form formula in Eq.~\eqref{eq:eigenvalues_real}.
    \item Compute $\lambda_{m,q}$ and $\hat{\lambda}_q = \max(\{\lambda_{m,q} : m \in [1, N-1]\})$ .
    \item Find the $q$-states for which $\hat{\lambda}_q < 0$; they are the stable ones \cite{sinha2025geometric}.
    \item For the $q$-states that are stable, compute the relative basin size from Eq.~\eqref{eq:basin_size}. 
    \item Predictions are compared with numerical simulations of Kuramoto networks given by Eq.~\eqref{eq:main_kuramoto}.
\end{enumerate}

\vspace{3.5cm}
%\subsection*{Acknowledgements}
\textbf{Acknowledgements:}
K.L.R. acknowledges support by the Max Planck Society.  A.M. acknowledges the financial support of the São Paulo Research Foundation (FAPESP), Grant No.~2023/08144-3. R.O.M.-T. acknowledges the São Paulo Research Foundation (FAPESP), Proc. No. 2015/50122-0 and Grant No. 2024/06718-5) and National Council for Scientific and Technological Development (CNPq), Proc. No. 408522/2023-2. R.C.B acknowledges the support of the Natural Sciences and Engineering Research Council of Canada (NSERC) [RGPIN-2026-05758, DGECR-2026-0006] and the Canada Research Chairs program.

\subsection*{Data and code availability}

An open-source repository with the codes and data used in this work is available at \href{https://github.com/budzinskilab/basin_kuramoto}{\textcolor{Cerulean}{github.com/budzinskilab/basin\_kuramoto}}.

%\bibliography{references.bib}


\end{document}